\documentclass[twocolumn,showpacs,preprintnumbers,floatfix,pre]{revtex4}
\usepackage{graphicx}
\usepackage{amsmath} 
\usepackage{color}

\begin{document}

\title{Toward Bernal Random Loose Packing through freeze-thaw cycling}

\author{F. Ludewig, N. Vandewalle, S. Dorbolo, M. Pakpour and G. Lumay}
\affiliation{GRASP, Universit\'e de Li\`ege, B-4000 Li\`ege, Belgium.}
\begin{abstract}
We study the effect of freeze-thaw cycling on the packing fraction of equal spheres immersed in water. The water located between the grains experiences a dilatation during freezing and a contraction during melting. After several cycles, the packing fraction converges to a particular value $\eta_{\infty} = 0.595$ independently of its initial value $\eta_0$. This behavior is well reproduced by numerical simulations. Moreover, the numerical results allow to analyze the packing structural configuration. With a Vorono\"i partition analysis, we show that the piles are fully random during the whole process and are characterized by two parameters: the average Vorono\"i volume $\mu_v$ (related to the packing fraction $\eta$) and the standard deviation $\sigma_v$ of Vorono\"i volumes. The freeze-thaw driving modify the volume standard deviation $\sigma_v$ to converge to a particular disordered state with a packing fraction corresponding to the Random Loose Packing fraction $\eta_{BRLP}$ obtained by Bernal during his pioneering experimental work. Therefore, freeze-thaw cycling is found to be a soft and spatially homogeneous driving method for disordered granular materials. 
\end{abstract}

\maketitle


How a large number of solid objects can fill a volume is one of the most puzzling problems in mathematics, science and engineering. This question concerns a broad range of systems: granular media, colloids, structures of living cells, amorphous solids, \textellipsis The relevant parameter that characterizes a pile of particles is the dimensionless packing fraction $\eta$, defined as the volume of all particles divided by the apparent volume of the assembly. This packing fraction has a maximum value $\eta_{fcc} = \pi / 3\sqrt{2} \simeq 0.74$ for identical spheres, corresponding to the face-centered cubic (fcc) lattice. The existence of this maximum was stated by Kepler in 1611 and demonstrated recently \cite{Hales2005}. The first systematic study of random equal spheres packings has been performed by Bernal {\it et al} \cite{Bernal1960}. The random packing of spheres was considered as a useful model for ideally simple liquid. The obtained range of random packing fraction was found to lie between two well-defined limits: the Bernal Random Loose Packing fraction $\eta_{BRLP} = 0.60$ and the Bernal Random Close Packing fraction $\eta_{BRCP} = 0.63$. The value of the packing fraction depends strongly on the history of the pile, i.e. on the way particles are sequentially placed in the assembly. More recently, the range of packing faction for random packings has been extended. By using fluidized bed techniques \cite{Swinner2005} or in a fluid providing a strong buoyancy \cite{Onoda1990}, one can decrease the random loose packing fraction to $\eta_{RLP} \simeq 0.55$. The present accepted value for the random close packing limit is around $\eta_{RCP} \simeq 0.64$ \cite{Makse2008}. The exact value of these limits and the link with the packing structure is still a matter of intense debate \cite{Aste2008,Tian2014,Klatt2014}. The situation is even more complex with non-spherical grains \cite{Chaikin,Behringer2014}. 

Different driving methods are leading to the modification of the packing fraction: tapping \cite{Nowak1998,Richard2005,Lumay2005}, cyclic shear \cite{Pouliquen2003}, flow-induced fluidization \cite{Swinney2005}, \textellipsis The existence of cohesive forces is also known to modify the packing fraction \cite{LumayNJP2007}. Recently, the effect of thermal cycling on packings has been studied experimentally \cite{Klein2006,Geminard2013} and numerically \cite{Taberlet2013}. Despite the low magnitude of the induced thermal expansion, a significative densification of the packing has been observed. 


Beyond the classical problem of packings, many phenomenons observed in nature are related to the particular behavior of wet granular materials submitted to temperature cycling: ice-lens formation in soil leading to frost heaving \cite{Peppin2011,Kurita2013}, landslides, structures formation in permafrost, stone heave and possibly some geological formations observed on Mars. Indeed, both granular materials and water have remarkable properties. The water in the liquid phase has a density maximum around $4^\circ$C and the solid phase density is lower than the liquid phase density. Moreover, comparing the other materials, water has a very high specific heat capacity, a high heat of vaporization, a high latent heat and a high surface tension. On the other hand, granular materials are out of equilibrium systems witch exhibits remarkable behaviors: phase segregation \cite{Kudrolli2004}, jamming transition \cite{Torquato2013}, extremely slow compaction dynamics \cite{Nowak1998,Richard2005,Lumay2005}, glassy dynamics \cite{Nagel2001}, \textellipsis Therefore, the association of water and granular material leads ineluctably to rich behaviors \cite{Xu2007,Fiscina2010}.

In the present letter, we show how the packing fraction of a spheres pile totally immersed in water is modified by successive freeze-thaw transitions. Moreover, the evolution of the packing structure during freeze-thaw transitions have been analyzed with Vorono\"i partition method applied on packings obtained by numerical simulations.


\begin{figure}
  \includegraphics[scale=0.3]{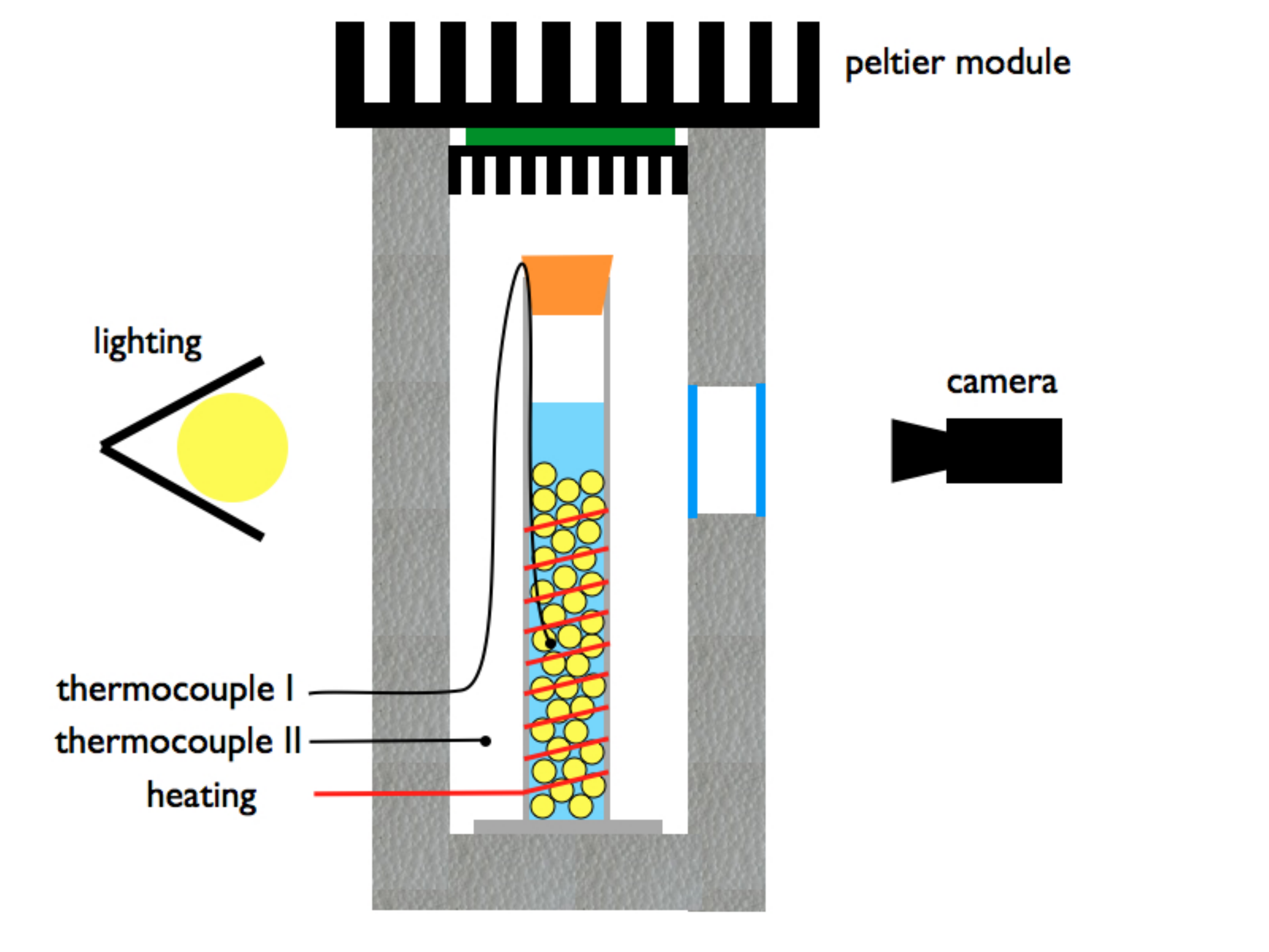}
  \caption{(Color online) Sketch of the experimental setup. A high CCD camera records the top of the packing placed in a glass tube. The pile is back-illuminated by a homogenous lighting system.}
\label{fig:SetUp}
\end{figure}

A sketch of the experimental set-up is presented in Figure \ref{fig:SetUp}. A borosilicate glass tube of internal diameter $D_{tube}=18.4$ mm is filled with water. Afterward, a granular material made of glass spheres with a diameter $D_{grain}=0.5$ mm is poured gently inside the tube. This method leads to a loose initial packing fraction. To obtain a higher value of the initial packing fraction, the tube is placed inside an ultrasonic cleaning tank during one minute. The vibrations induce a compaction of the granular bed. To perform temperature cycling, the tube is placed inside an isotherm box. The temperature is decreased with a Peltier module. The temperature increase is produced by a resistive wire winded around the glass tube. Two thermocouples measured respectively the temperature of the air inside the box and the temperature inside the granular material. A CCD camera takes pictures of the top of the granular pile. The position of the water meniscus is also visible on the pictures. The height of both water/air and granular/water interfaces are obtained by image treatment. Test runs with no thermal cycling were performed to ensure that vibrations do not affect the measurements. Moreover, thermal cycling have been performed on granular pile without water. In this dry case, we do not observe any significative variation of the packing fraction with our experimental set-up. 

The volume thermal expansion coefficient of the glass beads and of the tube are respectively $\beta_{glass} = 25.5 \; \mu K^{-1}$ and $\beta_{boro. \; glass} = 9.9 \; \mu K^{-1}$. The water thermal expansion coefficient varies significantly with the temperature and ranges between $-50 \; \mu K^{-1}$ to $207 \; \mu K^{-1}$ when the temperature goes from 1$^\circ$C to 20$^\circ$C. A rough estimation of the different material expansion for a temperature ramp between 0$^\circ$C and 20$^\circ$C gives 0.02\% for the tube, 0.05\% for the grains and 0.15\% for the water. Moreover, during the freezing transition, the water volume increase is around 8\%. These estimations show that the presence of water in the system is expected to influence deeply the thermal cycling compaction dynamics, in particular during freeze-thaw transitions. 

\begin{figure}
  \includegraphics[scale=0.4]{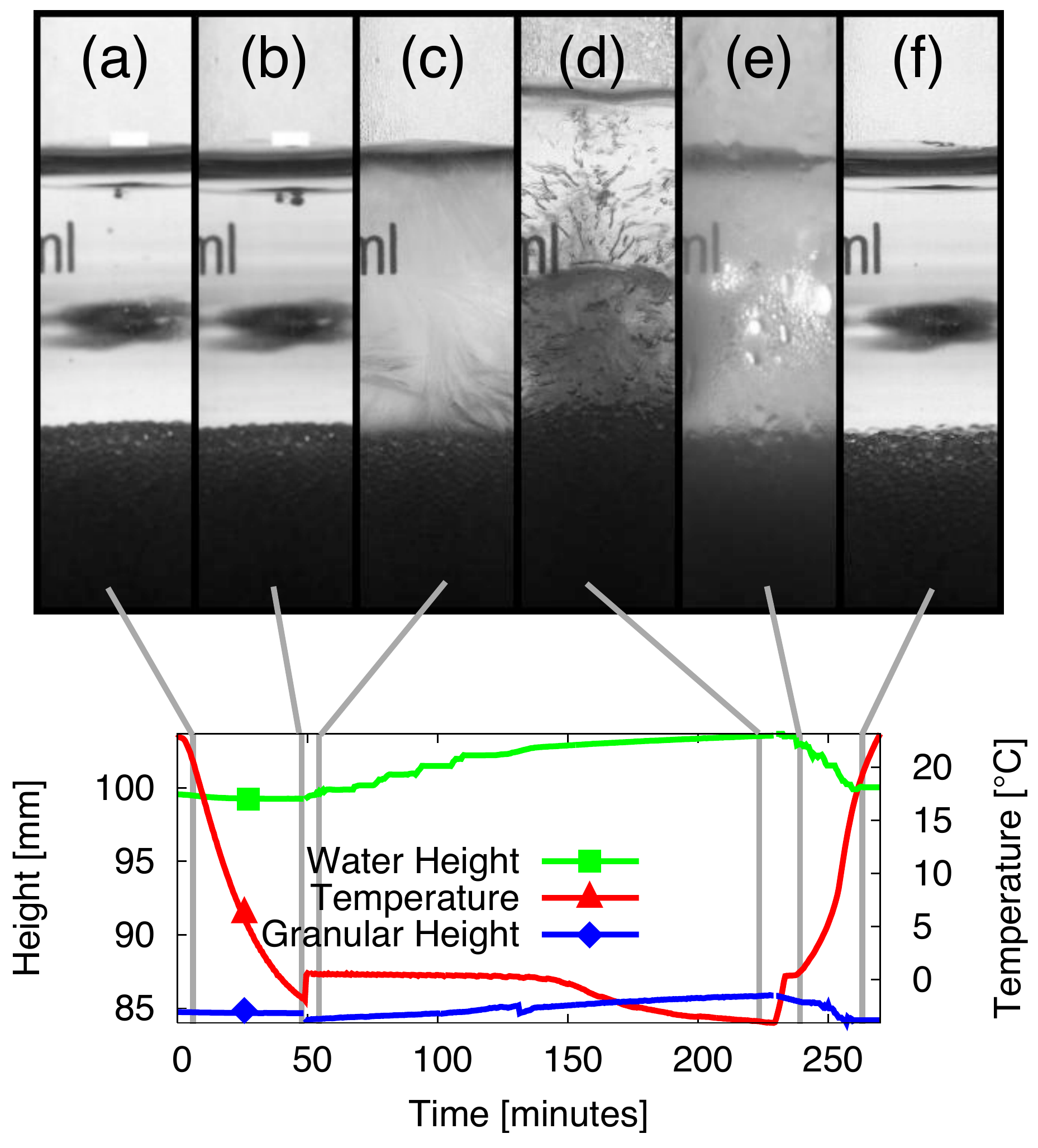}
  \caption{(Color online) Top: picture of the top of the pile (a) at the beginning of the cycle, (b) just before the freezing, (c) just after the freezing, (d) at the end of the freezing procedure, (e) during the heating and (f) at the end of the cycle. Bottom: temporal evolution of the temperature inside the pile, of the water height and of the granular height during one freeze-thaw cycle ($\eta_0 = 0.573$).}
  \label{fig:TemperatureCycles}
\end{figure}

The pile is submitted to freeze-thaw cycling with temperature ranging from $T_1 = -4^\circ C$ and $T_2 = 24^\circ C$ (see Figure \ref{fig:TemperatureCycles} (bottom)). The pile temperature reaches negative values before freezing. Therefore, a supercooling effect is observed. The time necessary to obtain a freezing of the system fluctuates from one cycle to the other. During the freezing transition, the temperature inside the pile goes quickly to $0^\circ$C. During this fast freezing step, the position of the granular/liquid and liquid/air interfaces do not change significantly (see Figure \ref{fig:TemperatureCycles} (Top)). However, during the temperature decrease after this freezing transition, both interfaces are going up. Then, a dilatation of the whole system is observed. Some steps are observed during this increase. When the pile temperature reaches $T_1= -4^\circ C$, the Peltier module is switched off and the heating is powered on. As a consequence, the temperature inside the pile increases to the temperature $T_2 = 24^\circ C$ with a plateau at $0^\circ$C. During the heating, the interfaces are going down. Some steps are also observed during this heating process.

\begin{figure}
  \includegraphics[scale=0.6]{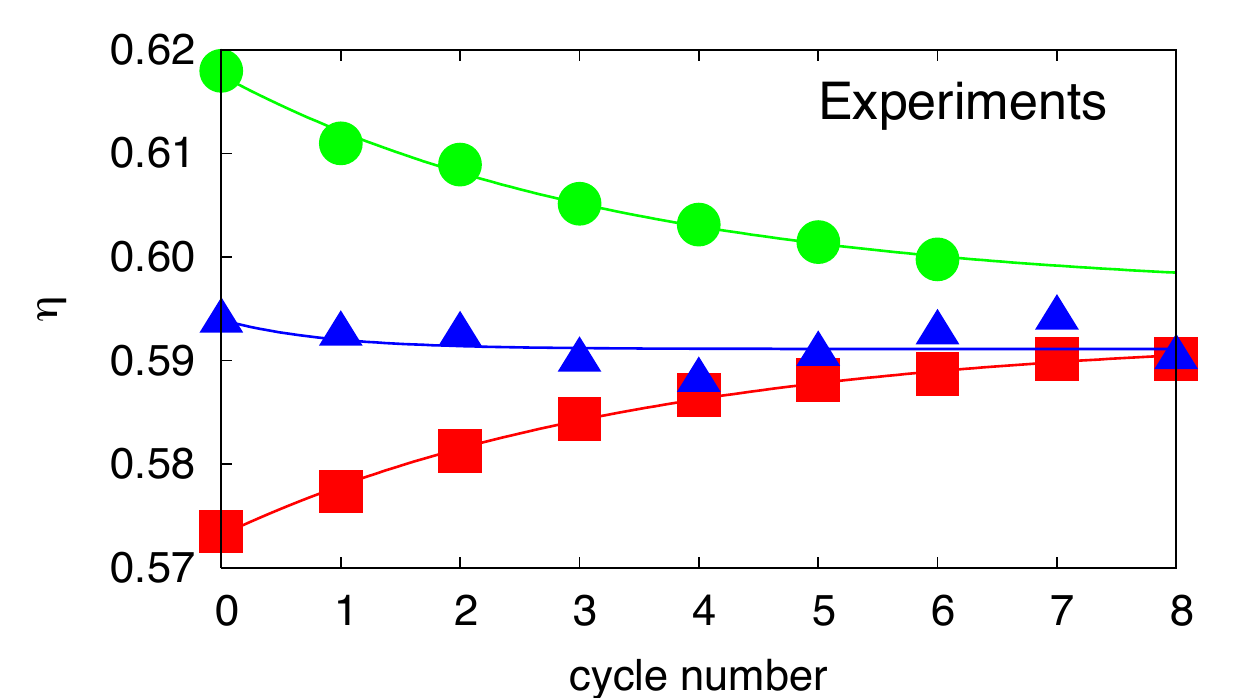}
  \includegraphics[scale=0.6]{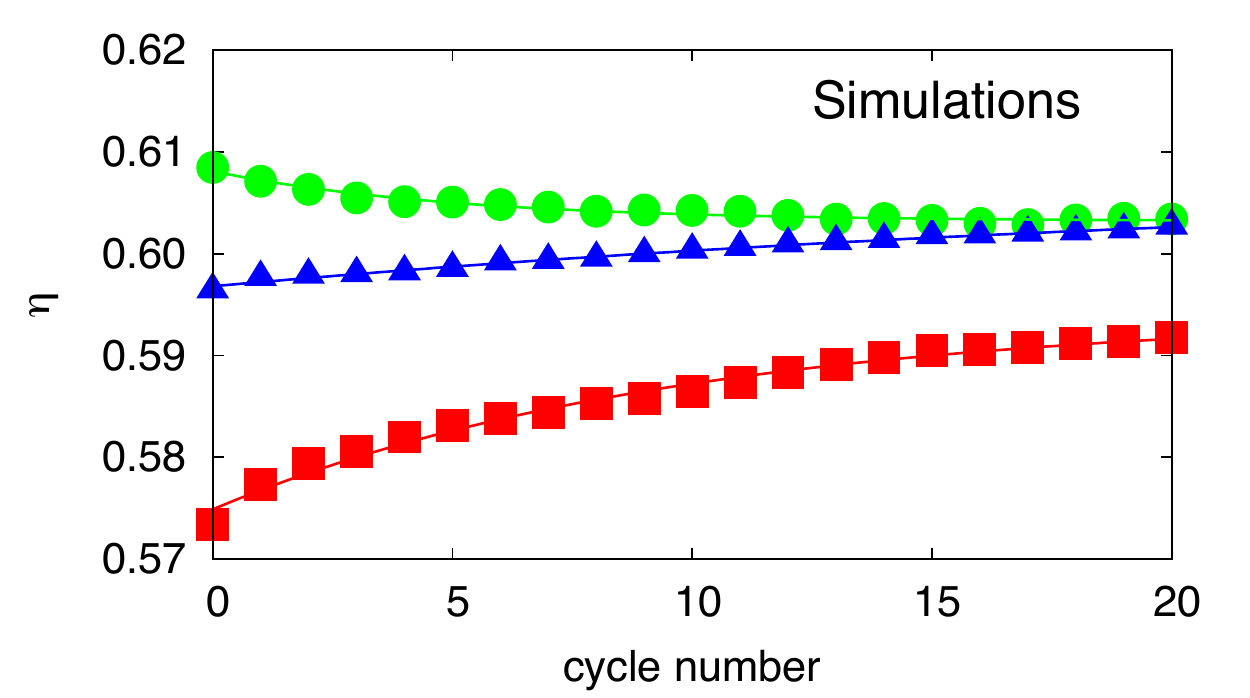}  
  \caption{(Color online) Evolution of the packing fraction $\eta$ as a function of the freeze-thaw cycle number for experiments (top) and simulations (bottom). The squares and the circles are corresponding respectively to a loose and a close initial packing. The triangles are corresponding to an initial packing fraction close to the Bernal Loose random Packing fraction $\eta_{BRLP} = 0.60$. The curves are fits from exponential law (see text).}
\label{fig:etaVScycles}
\end{figure}

The evolution of the packing fraction $\eta$ as a function of the cycle number $n$ is presented in Figure \ref{fig:etaVScycles} (top). Three experiments performed with different values of the initial packing fraction $\eta_0$ are presented. The experimental data are well fitted by the single exponential law proposed by Mehta {\it et al} \cite{Mehta1993} $\eta(n) = \eta_\infty - \Delta \eta e^{-n/\tau}$, where $\eta_\infty$, $\Delta \eta$ and $\tau$ are respectively the asymptotic packing fraction, the range of packing fractions and a characteristic cycle number. For low initial packing fraction, the freeze-thaw cycling induces a densification of the pile ($\eta_\infty = 0.592 \pm 0.001$, $\Delta \eta = 0.019 \pm 0.001$ and $\tau = 3.5 \pm 0.3$). On the other hand, a decompaction is observed for higher initial packing fraction ($\eta_\infty = 0.596 \pm 0.002$, $\Delta \eta = -0.021 \pm 0.001$ and $\tau = 3.4 \pm 0.7$). In both case, the packing fraction is found to converge to the Bernal Random Loose Packing fraction $\eta_{BRLP} = 0.60$. Moreover, the packing fraction is not influenced by freeze-thaw cycling when the initial packing fraction is close to $\eta_{BRLP}$. Therefore, this particular packing fraction should corresponds to a specific structural configuration of the packing.

\begin{figure}
  \includegraphics[scale=0.25]{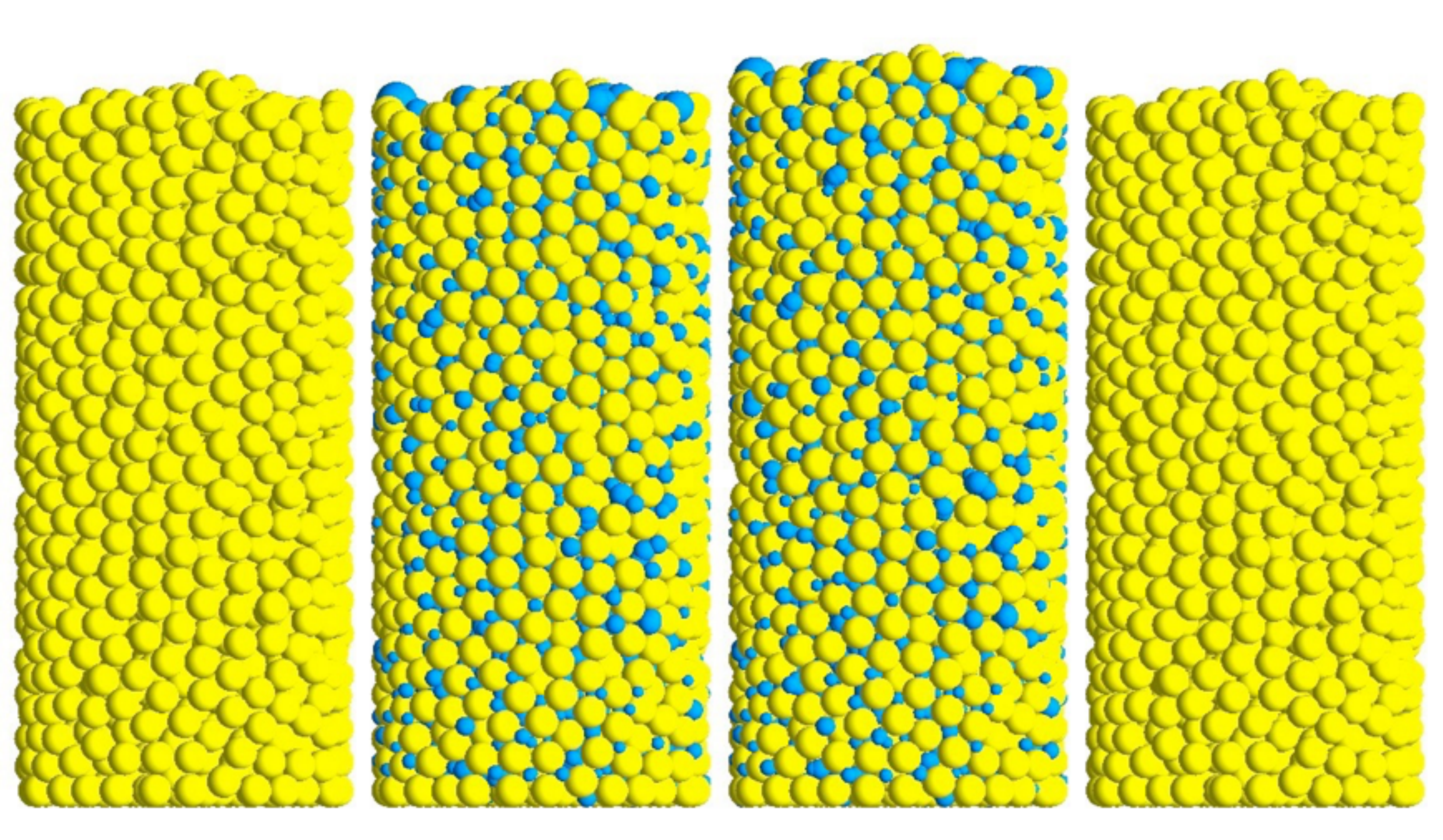}
  \caption{(Color online) Representation of the pile during one simulated freeze-thaw cycle. The grains are colored in yellow and the void spheres corresponding to the ice are colored in blue (color online). From the left to the right: initial pile with a low packing fraction, the void spheres are added, dilatation of the void spheres, final state with a higher packing fraction.}
\label{fig:FreezeThawCycleSimu}
\end{figure}

Although the compaction was expected for a loose initial packing fraction, the decompaction process of a dense pile is surprising. Moreover, the convergence of the packing fraction to the Bernal Random Loose Packing fraction $\eta_{BRLP} = 0.60$ is striking. In order to analyze the structural evolution of the packing during freeze-thaw cycling, numerical simulations have been performed. The model is based on molecular dynamics with tangential spring in order to produce a static pile. To obtain details about the numerical method, see ref \cite{Opsomer2013}. A tube of diameter $D_{tube}^{sim}= 10$mm is filled with N=1850 grains of diameter $D_{grain}^{sim} = 1$mm. The grains have the density of glass ($\rho = 2500 $ kg\;m$^{-3}$). After the initialization method, the voids between the grains are filled with spheres having the density of water. The void positions are obtained by Vorono\"i tessellation. Typically, 8000 void spheres are added. A freezing step consists in the dilatation of the void spheres, while a thaw step corresponds to a contraction of the void spheres. The evolution of the pile during a simulated freeze-thaw cycle is presented in Figure  \ref{fig:FreezeThawCycleSimu}. When two void spheres are overlapping, a spring link is defined between them in order to avoid collapse during the dilatation process. As shown by Figure \ref{fig:etaVScycles} (bottom), the compaction for loose initial packing fraction ($\eta_\infty = 0.594 \pm 0.001$, $\Delta \eta = 0.019 \pm 0.001$ and $\tau = 9.7 \pm 1.2$) and the decompaction for higher initial packing fraction ($\eta_\infty = 0.600 \pm 0.001$, $\Delta \eta = -0.005 \pm 0.001$ and $\tau = 5.0 \pm 0.5$) are well reproduced by the simulations. The characteristic cycle number obtained in simulations and in experiments are different because the size of the grains and of the container are different in experiments and in simulations.

\begin{figure}
  \includegraphics[scale=0.55]{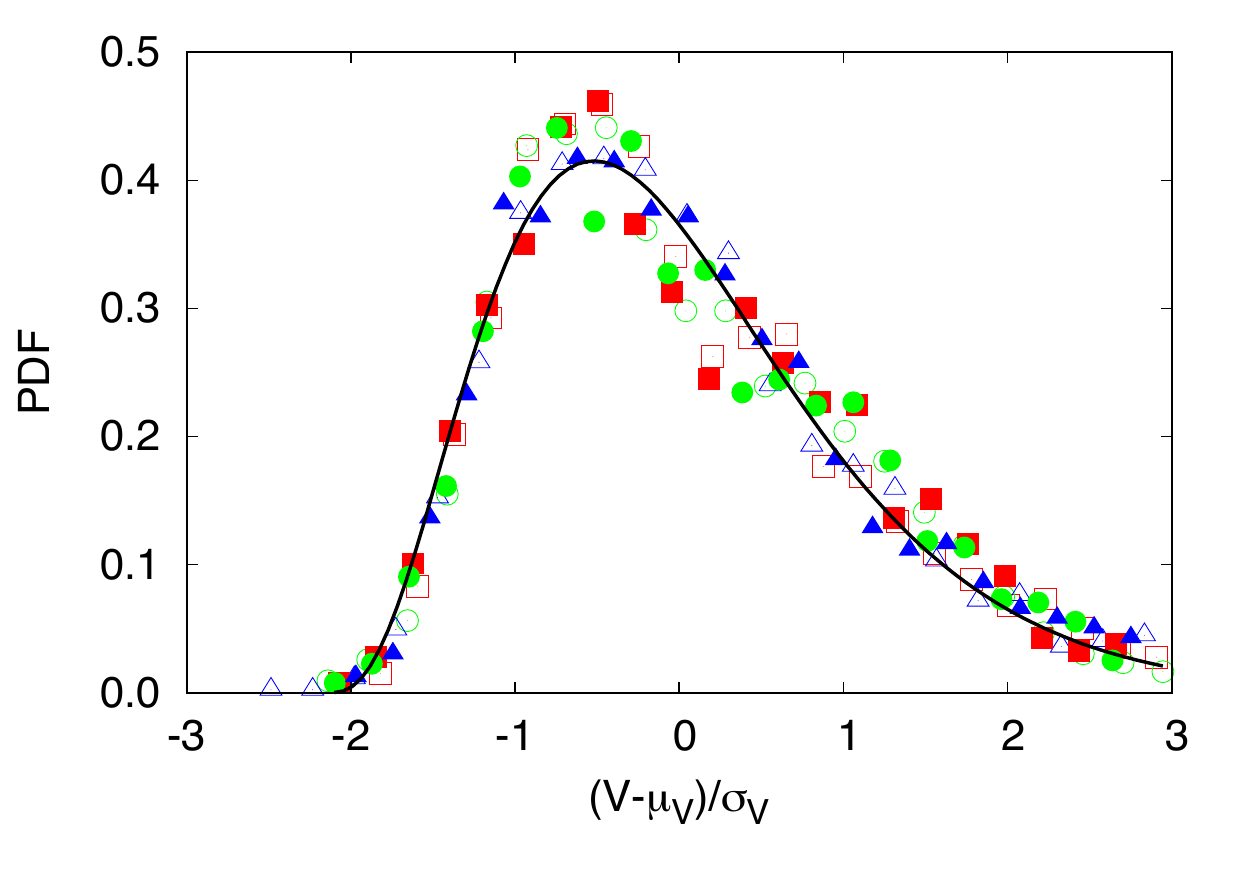}
  \caption{(Color online) Probability Density Function (PDF) of the Vorono\"{i} volumes $V$ for the initial piles (open symbols) and after 20 freeze-thaw cycling (plain symbols). The volumes $V$ are normalized with the volume average $\mu_v$ and the volume standard deviation $\sigma_v$. The convention for the symbol shapes are the same than in Figure \ref{fig:etaVScycles}. The data have been obtained from numerical simulations. The line corresponds to the Gamma function.}
\label{fig:DistriVoronoi}
\end{figure}

The packings obtained by numerical simulations have been analyzed with Vorono\"i partition. The probability density functions of the Vorono\"{i} cell volumes $V$ for all the initial piles and for all the piles after 20 freeze-thaw cyclings are presented in Figure \ref{fig:DistriVoronoi}. The volume normalization $(V-\mu_v)/\sigma_v$ with the Vorono\"{i} volume average $\mu_v$ and the Vorono\"{i} volume standard deviation $\sigma_v$ induces a collapsing of the density functions. Then, the volume average $\mu_v$ and the volume standard deviation $\sigma_v$ are the main parameters characterizing the system. The volume average $\mu_v$ is related to the packing fraction $\eta = V_{grain}/\mu_v$, where $V_{grain}$ is the volume of one grain. Moreover, the Gamma shape of the density functions is characteristic of a fully random system \cite{Tanemura2003}. We have tested with numerical simulations that an other driving mechanism like vibrations induces a deviation from the Gamma shape due to the apparition of ordered domains. Therefore, freeze thaw cycling is found to be a soft driving method for disordered granular materials. The driving modify the volume standard deviation $\sigma_v$ to converge to a particular disordered state with a packing fraction close to $\eta_{BRLP} = 0.60$. Contrary to driving processes based on mechanical agitation, we do not observe any convective motions, any grains spatial organization in the bulk and any grains ordering close to the container wall \cite{Nowak1998,Richard2005,Lumay2005}. Moreover, the driving method based on freeze-thaw cycling is spatially homogenous.

\begin{figure}
  \includegraphics[scale=0.22]{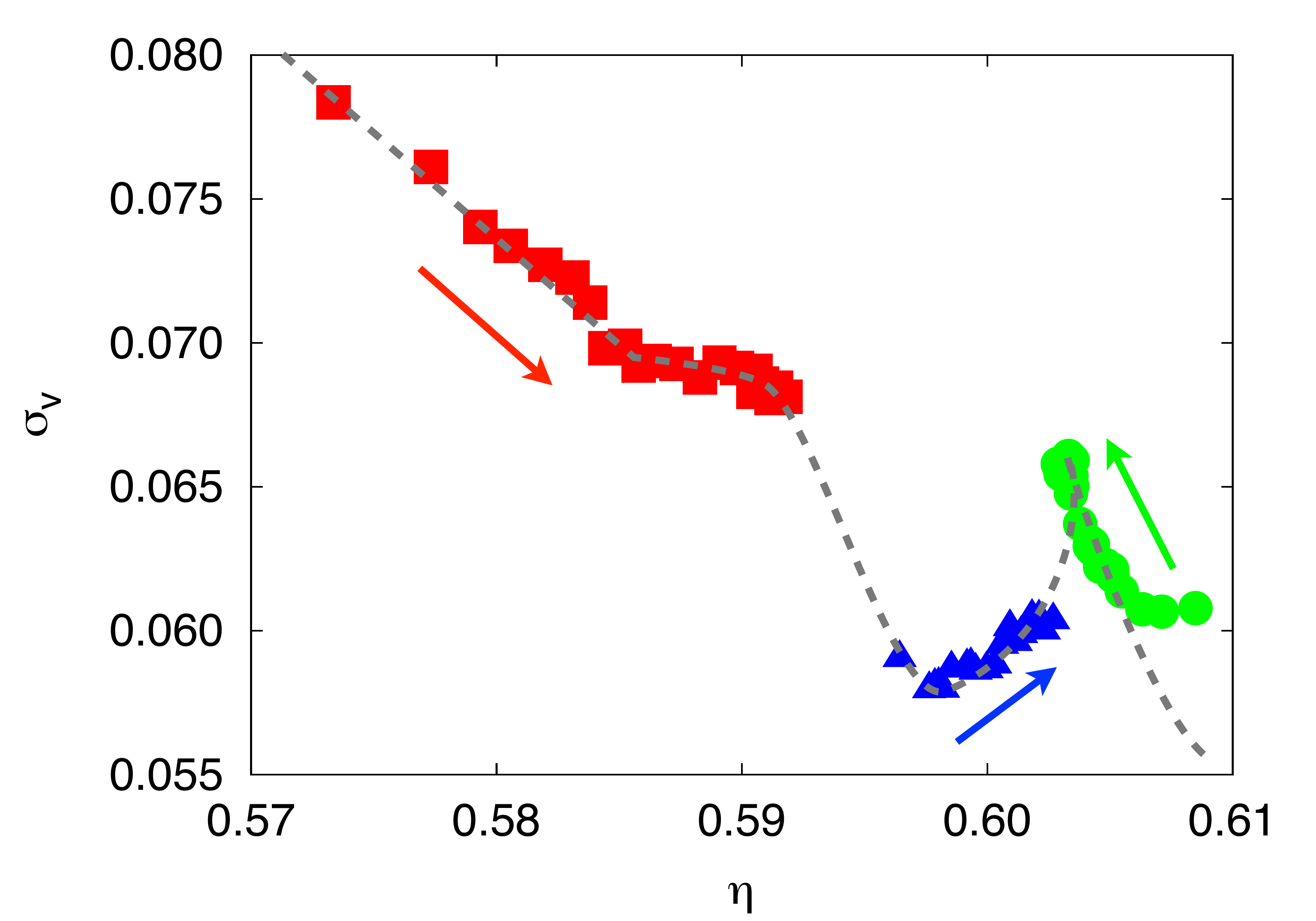}
  \caption{(Color online) Evolution of the Vorono\"{i} volume standard deviation $\sigma_v$ as a function of the packing fraction $\eta$ during 20 cycles for three different initial packing fractions. The data have been obtained from numerical simulations. The dashed curves are guide for the eyes inspired from the work of Aste {\it et al.} \cite{Aste2008}.}
\label{fig:STDdistriVoronoi}
\end{figure}

A diagram with the main parameters obtained from the simulated packings (packing fraction $\eta= V_{grain}/\mu_v$ and standard deviation $\sigma_v$ of the Vorono\"i volumes) is presented in Figure \ref{fig:STDdistriVoronoi}. Aste {\it et al.} \cite{Aste2008} have shown that $\sigma_v$ decreases between the random loose and the random close packing limits. Moreover, they observed a small but sizable local minimum around $\eta = 0.60$. This behavior was observed with experimental and numerical granular assemblies created in different conditions. The freeze-thaw cycling allow to move in this diagram along the master curve describe by Aste {\it et al.}. We have checked numerically that a stronger mechanical driving mechanism like vibrations induces a deviation from this master curve due to the apparition of ordered clusters in the packing and close to the wall. 

In summary, the evolution of the packing fraction of an assembly of spheres immersed in water and submitted to freeze-thaw cycling has been investigated. As already observed in the dry case, the packing fraction increases when starting from a loose configuration. However, when starting from a close configuration, the freeze-thaw cycling induces a decompaction of the pile. Independently of the initial packing fraction $\eta_0$, the packing fraction converges to a particular value $\eta_{\infty} = 0.595$. This final packing fraction is equivalent to the Random Close Packing fraction obtained by Bernal during his precursor experimental works performed in the sixties. Contrary to driving processes based on mechanical agitation, we do not observe any convective motions, any grains spatial organization in the bulk and any grains ordering close to the container wall. In addition, the driving based on freeze-thaw cycling is spatially homogenous. A structural analysis based on Vorono\"i partition has been performed with numerical simulations. The distribution of the Vorono\"i has shown that the packings are fully random during the whole process. Moreover, the packing fraction $\eta= V_{grain}/\mu_v$ and the Vorono\"i volume standard deviation $\sigma_v$ are the main parameters characterizing the piles.


This work was financially supported by the FNRS (Grant PDR T.0043.14) and by the University of Li\`ege (Starting Grant C-13/88). FL thank the Shape-SPS project (Wallonia) for supporting the development of the numerical model.


\begin{thebibliography}{2} 
\bibitem{Hales2005} T. Hales, Ann. of Math. {\bf 162}, 1065–1185 (2005)
\bibitem{Bernal1960} J.D. Bernal and J. Mason, Nature {\bf 188}, 910 (1960)
\bibitem{Swinner2005} Matthias Schr\"{o}ter, D. I. Goldman, and H. L. Swinney, Phys. Rev. E  {\bf 71}, 030301(R) (2005)

\bibitem{Onoda1990} G. Y. Onoda  and E. G. Liniger , Phys. Rev. Lett. {\bf 64}, 2727 (1990)
\bibitem{Makse2008} C. Song, P. Wang, and H. A. Makse, Nature {\bf 453}, 29 (2008)
\bibitem{Aste2008} T. Aste and T. Di Matteo, Eur. Phys. J. B {\bf 64}, 511 (2008)
\bibitem{Tian2014} Z. A. Tian, K. J. Dong, and A. B. Yu, Phys. Rev. E {\bf 89}, 032202 (2014)
\bibitem{Klatt2014} M. A. Klatt, and S. Torquato, Phys. Rev. E {\bf 90}, 052120 (2014)

\bibitem{Chaikin} A. Donev, I. Cisse, D. Sachs, E. A. Variano, F. H. Stillinger, R. Connelly, S. Torquato, and P. M. Chaikin, Science {\bf 303}, 990-993 (2004)

\bibitem{Behringer2014} S. Farhadi and R.P. Behringer, Phys. Rev. Lett. {\bf 112}, 148301 (2014)

\bibitem{Nowak1998} E. R. Nowak, J. B. Knight, E. Ben-Naim, H. M. Jaeger, and S. R. Nagel, Phys. Rev. E \textbf{57}, 1971 (1998)

\bibitem{Richard2005} P. Richard, M. Nicodemi, R. Delannay, P. Ribi\`ere, and D. Bideau, Nature Materials \textbf{4}, 121 (2005)

\bibitem{Lumay2005} G. Lumay and N. Vandewalle, Phys. Rev. Lett. \textbf{95}, 028002 (2005)

\bibitem{Pouliquen2003} O. Pouliquen, M. Belzons and M. Nicolas, Phys. Rev. Lett. \textbf{91}, 014301 (2003)

\bibitem{Swinney2005} M. Schroter and D.I. Goldman and H.L. Swinney, Phys. Rev. E \textbf{71}, 030301(R) (2005)

\bibitem{LumayNJP2007} G. Lumay and N. Vandewalle, New J. of Phys. {\bf 9}, 406 (2007).

\bibitem{Klein2006} K. Chen, J. Cole, C. Conger, J. Draskovic, M. Lohr, K. Klein, T. Scheidemantel and P. Schiffer, Nature \textbf{442}, 257 (2006)
\bibitem{Geminard2013} B. Blanc and J.-C. G\'eminard, Phys. Rev. E {\bf 88}, 022201 (2013)
\bibitem{Taberlet2013} B. Percier, T. Divoux and N. Taberlet, Europhys. Lett. {\bf 104} 24001(2013)
\bibitem{Peppin2011} R.W. Style, S. S. L. Peppin, A. C. F. Cocks, and J. S. Wettlaufer, Phys. Rev. E {\bf 84}, 041402 (2011)

\bibitem{Kurita2013} T. Saruya, K. Kurita and A. W. Rempel, Phys. Rev. E 87, 032404 (2013)

\bibitem{Kudrolli2004} A. Kudrolli, Rep. Prog. Phys. \textbf{67}, 209 (2004)

\bibitem{Torquato2013}  S. Atkinson, F. H. Stillinger, and S. Torquato, Phys. Rev. E {\bf 88}, 062208 (2013)

\bibitem{Xu2007} Q. Xu, A. V. Orpe, and A. Kudrolli, Phys. Rev. E {\bf 76}, 031302 (2007)

\bibitem{Fiscina2010} J. E. Fiscina, G. Lumay, F. Ludewig, and N. Vandewalle, Phys. Rev. Lett. {\bf 105}, 048001 (2010)

\bibitem{Nagel2001} C. S. O’Hern, S. A. Langer, A. J. Liu, and S. R. Nagel, Phys. Rev. Lett. {\bf 86}, 000111 (2001)

\bibitem{Mehta1993}G. C. Barker and Anita Mehta, Phys. Rev. E {\bf 47}, 184 (1993)

\bibitem{Opsomer2013} E. Opsomer, M. Noirhomme, N. Vandewalle, F. Ludewig, Phys. Rev. E  {\bf 88}, 012202 (2013)

\bibitem{Tanemura2003} M. Tanemura, Forma {\bf 18}, 221–247 (2003)

\end{thebibliography}
\end{document}